\documentclass[a4paper,aps,amsmath,amssymb,showpacs,10pt,twocolumn]{revtex4}   
\usepackage{epsfig}   
\usepackage{graphicx}   
\usepackage{amsmath,amssymb}

\newcommand{\ket}[1]{\left|#1\right\rangle}   
\newcommand{\bra}[1]{\left\langle #1\right|}   
\newcommand{\bracket}[1]{\left\langle #1\right\rangle}   
   
\newcommand{\be}{\begin{equation}}   
\newcommand{\ee}{\end{equation}}   
\newcommand{\bd}{\begin{displaymath}}   
\newcommand{\ed}{\end{displaymath}}

\newcommand{\bsigma}{{\mbox{\boldmath $\sigma$}}}

\newcommand{\bP}{\ensuremath{\mathbf{P}}}

\newcommand{\bS}{\ensuremath{\mathbf{S}}}

\begin{document}   
\title{On exact mappings between fermionic Ising spin glass and classical spin glass models}   
\author{Isaac P\'erez Castillo}   
\author{David Sherrington}   
\affiliation{Rudolf Peierls Center for Theoretical Physics, 
University of Oxford, 1 Keble Road, Oxford, OX1 3NP, UK}   
\begin{abstract}   
We present in this paper exact analytical expressions for the   
thermodynamical properties and Green functions of a certain family   
of fermionic Ising spin-glass models with Hubbard interaction, by   
noticing that their Hamiltonian is a function of the number operator   
only. The thermodynamical properties are mapped to the classical   
Ghatak-Sherrington spin-glass model while the the Density of States   
(DoS) is related to its joint spin-field distribution. We discuss the   
presence of the pseudogap in the DoS with the help of this mapping.  
\end{abstract}   
\maketitle   
\section{Introduction}   
While ferromagnetism is theoretically grounded in models lying between  
two extreme pictures, that of localised-spins and that of  
itinerant-electron theory, the theoretical description of spin-glass  
systems has been focused  mainly on models of localised-spins, whose  
paradigms are the Edwards-Anderson (EA) and Sherrington-Kirkpatrick  
(SK) models \cite{Edwards1975,Sherrington1975}. One reason  
for this is that experimentally most of the classic magnetic materials  
presenting spin-glass behaviour correspond to this description of  
localised spins, another is that they encapsulate many aspects of  
theoretical challenge, while a third is that they relate to or emulate  
many problems of wider interest in the statistical physics of complex  
systems.   
  
There do exist, however, spin glass materials which are more  
appropriately described in terms of itinerant electrons, requiring  
models that treat magnetic and conducting properties on the same  
footing.  While early models can be found in  
\cite{Sherrington1974,Hertz1979} and the full problem remains to be  
tackled, our main goal here is to discuss a restricted class of models  
introduced first by Oppermann \textit{et al.} \cite{Oppermann1993}, and  
normally referred as \textit{fermionic Ising spin-glass models}  
\cite{Rosenow1996,Oppermann1998,Oppermann1998a,Oppermann1999,  
Rehker1999,Feldmann2000a,Feldmann2000,Oppermann2003}.   
  
Even within an SK-like (infinite-ranged exchange) itinerant fermionic 
model it is a very significant challenge to  
treat conducting and magnetic properties together. Hence, as a first  
step towards their understanding, simplified models have been studied  
in the so-called \textit{insulating limit}  
\cite{Rosenow1996,Oppermann1998,Oppermann1998a,Oppermann1999,  
Rehker1999,Feldmann2000a,Feldmann2000,Oppermann2003}. This 
removes the  \textit{essentially quantum complexity} of the model 
and allows a classical treatment,  
albeit still with interesting consequences. 
  
Independently of whether the insulating limit of fermionic Ising  
spin-glass models may or may not be useful to better understand real  
itinerant spin glasses, it is clear that, at least, we must fully  
understand the models arising from this limit. 
To this end, our goal  
is to point out that \textit{not only} are the fermionic Ising  
spin-glass models completely mappable to classical spin-glass models  
at the level of the thermodynamics\cite{Feldmann2000a}, \textit{but also}  
the densities of states (DoS), and hence the local (quantum) Green functions,
have a classical derivation and indeed are 
given by distributions of local fields of a corresponding  
classical model, without the need for sophisticated quantum treatment.

Hence, we can exploit all the knowledge of classical  
spin glasses to shed light on the fermionic Ising spin-glass models in  
the insulating limit. In particular, the existence of a pseudogap in  
the DoS at half-filling and without a Hubbard term emerges as an
immediate consequence of the mapping, when account is taken of the 
well-known fact that at zero temperature the local field 
distribution of the SK spin glass has such a pseudogap.

In turn, this implies that the observed \textit{strong corrections} to
the DoS due to steps in the replica symmetry breaking
\cite{Oppermann2003} do not have a fundamentally quantum origin.

We also notice, \textit{en passant}, the strong temperature dependence
of the DoS, mirroring that of the field distribution of the SK model.

This paper is organised as follows: in section
\ref{sec:modeldefinitions} the fermionic Ising spin-glass model is
presented and some limits as a function of its parameters
discussed. In section \ref{sec:mapping} we map this model to the
Ghatak-Sherrington model and express the DoS as a function of its
joint spin-field distribution. Then in section \ref{sec:physical} we
discuss the mappings from a physical perspective and in section
\ref{sec:pseudogap} we discuss the existence of a DoS pseudogap in the
light of the mapping.  The last section of the main text is for the
conclusions.
   
\section{Model definitions}   
\label{sec:modeldefinitions}   
Our starting point is the following model for itinerant electrons   
involving frustrated magnetic order   
\begin{equation}   
\begin{split}   
\widehat{\mathcal{H}}&=U\sum_{i=1}^N   
\widehat{n}_{i\uparrow}\widehat{n}_{i\downarrow}-\sum_{i<j=1}^N   
\mathcal{J}_{ij}\widehat{\sigma}^z_i\widehat{\sigma}^z_j   
-\mu\sum_{i=1}^N\sum_{s\in\{\uparrow,\downarrow\}}\widehat{n}_{is}\\   
&+\widehat{\mathcal{H}}_{\text{rest}}   
\label{eq:quantum_Hamiltonian}   
\end{split}   
\end{equation}   
where the couplings $\mathcal{J}_{ij}$ are   
drawn randomly and independently from a distribution   
\begin{equation}   
\begin{split}   
P(\mathcal{J}_{ij})=\frac{1}{\sqrt{2\pi\mathcal{J}^2/N}}\exp\left[-\frac{N}  
{2\mathcal{J}^2}\left(\mathcal{J}_{ij}-\frac{\mathcal{J}_0}{N}\right)^2\right].   
\label{eq:distribution_J}   
\end{split}   
\end{equation}   
The spin and charge operators are defined by   
\begin{equation}   
\widehat{\sigma}^z_{i}=\widehat{n}_{i\uparrow}-\widehat{n}_{i\downarrow},\quad  
\widehat{n}_{is}=\widehat{a}^\dag_{is} \widehat{a}_{is} \,,   
\end{equation}   
where $\widehat{a}^\dag_{is}$ and $\widehat{a}_{is}$ are respectively  
the fermion creation and annihilation operators.  The label  
$s\in\{\downarrow,\uparrow\}\equiv\{-1,1\}$ indicates the spin state.  
The Hamiltonian $\widehat{\mathcal{H}}_{\text{rest}}$ contains those  
terms that can not be expressed as a function of the number operator  
only; as, for example, the hopping and the pair-hopping terms,  
\textit{viz.}  
\begin{equation}   
\begin{split}   
\widehat{\mathcal{H}}_{\text{hopping}}&=\sum_{(i,j)}\sum_{s\in\{\uparrow,\downarrow\}}t_{ij} 
\widehat{a}_{is}^\dag\widehat{a}_{js},\\  
\widehat{\mathcal{H}}_{\text{hopping}}^{\text{pair}}&=\sum_{(i,j)}t^{\text{pair}}_{ij} 
\widehat{a}_{i\downarrow}^\dag\widehat{a}_{i\uparrow}^\dag\widehat{a}_{j\uparrow}\widehat{a}_{j\downarrow}\ ,  
\end{split}   
\end{equation}  
as well as transverse spin-exchange terms. 
  
The class of models described by \eqref{eq:quantum_Hamiltonian} with  
$\widehat{\mathcal{H}}_{\text{rest}}=0$ have been called fermionic  
Ising spin-glass models (FISG) \cite{Oppermann1999a} and may be considered as  
the insulating limit of the larger class of itinerant models described  
by the Hamiltonian \eqref{eq:quantum_Hamiltonian}. Henceforth, we  
consider this limit.  
  
In the past these models have been studied using techniques of  
coherent fermionic states (see for instance \cite{Oppermann2003}). 
Our purpose here is to  
point out that quantum techniques are unnecessary and a classical  
treatment suffices.

 \section{Mapping to classical spin-glass models}   
\label{sec:mapping}   
For a general fermionic problem, the coherent states representation is
a powerful technique and is likely to be useful for a treatment of the
full quantum Hamiltonian \eqref{eq:quantum_Hamiltonian}. However, for
the unfamiliar its use is likely to obscure a simplicity of the
insulating case. Hence, here we proceed differently, in what we
consider to be a much simpler way, for the Hamiltonian
 
\begin{equation}   
\begin{split}   
\widehat{\mathcal{H}}&=U\sum_{i=1}^N   
\widehat{n}_{i\uparrow}\widehat{n}_{i\downarrow}-\sum_{i<j=1}^N   
\mathcal{J}_{ij}\widehat{{\sigma}_i}^z\widehat{{\sigma}_j}^z   
-\mu\sum_{i=1}^N\sum_{s\in\{\uparrow,\downarrow\}}\widehat{n}_{is}.  
\label{eq:classical_Hamiltonian}   
\end{split}   
\end{equation}  
First we note that the Hamiltonian is function of the number operator  
$\widehat{n}_{is}$ only. We are therefore in the ideal position of  
knowing the eigenstates of the Hamiltonian exactly. If we define  
\begin{equation}   
\ket{\textbf{n}}=\prod_{i=1}^N\ket{n_{i\uparrow}n_{i\downarrow}}_{i}\,,   
\end{equation}   
then the partition function expressed in this set of states becomes   
\begin{equation}   
\mathcal{Z}(\beta)=\sum_{\textbf{n}}e^{-\beta\mathcal{H}(\textbf{n})}\,,   
\end{equation}   
where $\mathcal{H}(\textbf{n})$ is the Hamiltonian  
\eqref{eq:classical_Hamiltonian} with the operators now just  
numbers. This Hamiltonian is quite similar to such of a 3-state spin glass  
model. In order to make this similarity more apparent we express the  
partition function and the Hamiltonian as a function of the two new  
variables $S_{i}=n_{i\uparrow}-n_{i\downarrow}$ and  
$\tau_{i}=n_{i\uparrow}+n_{i\downarrow}$. After doing the trace with  
respect to the variables $\tau_i$, we end up with the partition  
function  
\begin{widetext}   
\begin{equation}   
\begin{split}   
\mathcal{Z}(\beta)=\sum_{\bS}e^{-\beta\mathcal{H}_{\text{GS}}(\bS)},  
\quad \mathcal{H}_{\text{GS}}(\bS)=-\sum_{i<j=1}^N \mathcal{J}_{ij}  
S_iS_j-D\sum_{i=1}^NS_i^2-\frac{N}  
{\beta}\log\Big(1+e^{-\beta U+2\beta\mu}\Big)\,,   
\label{eq:GS}   
\end{split}   
\end{equation}   
\end{widetext}   
with    
\begin{equation}   
D=\mu-T\log\Big(1+e^{-\beta U+2\beta\mu}\Big)   
\end{equation}   
and with notation $\bS=(S_1,\ldots,S_N)$, $S_i\in\{0,\pm1\}$. The  
Hamiltonian \eqref{eq:GS} is known in classical spin glass literature  
as the Ghatak-Sherrington (GS) model \cite{Ghatak1977}, a particular  
case of the Blume-Emery-Griffiths-Capel spin-glass model  
\cite{Crisanti2004} without biquadratic interaction.  
 
In the GS formulation, the new spin variables $\bS$ do not reflect the  
difference between unoccupied and doubly occupied sites. It is,  
therefore, useful to consider the calculation of the expectation  
value of the fermion number  
operator in this formulation. By defining  
\begin{equation}  
n=\frac{1}{N}\sum_{i=1}^{N}\sum_{s=\pm1}\bracket{\widehat{n}_{is}}_{\mathcal{H}_\text{FISG}},\quad  
\rho=\frac{1}{N}\sum_{i=1}^N\bracket{S_i^2}_{\mathcal{H}_\text{GS}}\,,  
\end{equation}  
then we can write the following relation  
\begin{equation}  
n=\frac{2(1-\rho)}{e^{\beta(U-2\mu)}+1}+\rho\,,  
\label{eq:n_rho}  
\end{equation}  
with $\bracket{\,\,\bullet\,\,}_{\mathcal{H}}$ the thermal average  
with respect a Hamiltonian $\mathcal{H}$, \textit{viz.}  
\begin{equation}   
\bracket{\cdots}=\mathcal{Z}^{-1}(\beta)\text{Tr}e^{-\beta\widehat{\mathcal{H}}}(\cdots)\,,   
\end{equation}   
where $\text{Tr}$ denotes the trace.  Thus, half-filling ($n=1$)  
corresponds to $\mu=U/2$.  
  
This mapping between the fermionic Ising spin-glass model
\eqref{eq:quantum_Hamiltonian} and the classical SG model
\eqref{eq:GS} was already noticed and used fruitfully in
\cite{Feldmann2000a}, but unfortunately these authors appear not to
have noticed that the Green function, and the DoS which can be derived
from it, can also be obtained simply from the classical
model\footnote{In later work \cite{Oppermann2003} one of these
authors, together with one of the present authors, has demonstrated
that within a replica treatment of the coherent state formulation the
FISG requires only static terms and examination shows their
consequence to be the same as would be obtained from a replica
treatment of the classical model, but this point was not made
explicitly, as neither was the correspondence even before replication
that we demonstrate in the present paper.}.

 Indeed, instead of using the  
fermionic path integral definition for the DoS, let us start with the  
standard definition for the retarded Green function  
\begin{equation}   
\mathcal{G}_{ij}^{s\,s'}(t-t')=-i\theta(t-t')\bracket{\{\widehat{a}_{is}(t),\widehat{a}^\dag_{js'}(t')\}}\,,   
\label{eq:Greenfunction_R}   
\end{equation}   
with $\{\widehat{A},\widehat{B}\}\equiv  
\widehat{A}\widehat{B}+\widehat{B}\widehat{A}$ and the creation and  
annihilation operators given in the Heisenberg representation  
\begin{equation}   
\begin{split}   
\widehat{a}_{is}(t)=e^{\frac{i}{\hbar} \widehat{\mathcal{H}}t}\widehat{a}_{is}e^{-\frac{i}{\hbar} \widehat{\mathcal{H}}t}\,.   
\end{split}   
\end{equation}   
Using the set of states $\ket{\textbf{n}}$ and after some standard  
manipulations the retarded Green function takes the form  
\begin{equation}   
\mathcal{G}_{ij}^{s\,s'}(t-t')=\delta_{i,j}\delta_{s,s'}\int_{-\infty}^{\infty} \frac{d\omega}{2\pi }e^{-i \omega (t-t')} \mathcal{G}_{i}^{s}(\omega)\,,   
\end{equation}   
with   
\begin{equation}   
\begin{split}   
\mathcal{G}_{i}^{s}(\omega)&=\mathcal{Z}^{-1}(\beta)\sum_{\textbf{n},\textbf{m}}|\bra{\textbf{n}}\widehat{a}_{i  
s}\ket{\textbf{m}}|^2\\  
&\times\frac{e^{-\beta\mathcal{H}(\textbf{n})}+e^{-\beta\mathcal{H}(\textbf{m})}}{\omega-[\mathcal{H}(\textbf{m})-\mathcal{H}(\textbf{n})]/\hbar+i\delta}\,.  
\label{eq:greenfunction123}   
\end{split}   
\end{equation}   
Notice that the above expression is fully general for any Hamiltonian  
system depending on the number operator only. Even though we could  
continue with the general calculation quite easily, at this stage we  
believe it to be helpful first to analyse the family of fermionic  
Ising spin-glass models without Hubbard interaction, which have been  
studied extensively with coherent-state methods  
\cite{Feldmann2000a,Feldmann2000,Oppermann1999,Rehker1999}.  
\subsection{Fermion Ising spin-glass model without Hubbard interaction}   
Due to the presence of $|\bra{\textbf{n}}\widehat{a}_{i  
s}\ket{\textbf{m}}|^2$ in eq. \eqref{eq:greenfunction123} the only  
states which contribute have  
\begin{equation}   
\begin{split}   
\mathcal{H}(\textbf{m})-\mathcal{H}(\textbf{n})&=-sh_{i}(\bS)-\mu,   
\end{split}   
\end{equation}   
with $h_{i}(\bS)=\sum_{j(\neq i)}^N \mathcal{J}_{ij}S_j$, the local  
field at site $i$. Hence, after some algebra, we can write  
\begin{equation}   
\begin{split}   
\mathcal{G}_{i}^{s}(\omega)&=\int \, \frac{d h}{\omega +(\mu+sh )/\hbar+i\delta}\\   
&\times\mathcal{Z}^{-1}(\beta)\sum_{\textbf{n}}e^{-\beta\mathcal{H}(\textbf{n})}\delta\left[h-h_{i}(\bS)\right]\,.   
\end{split}   
\end{equation}   
Notice that all the spin dependence is in the last term. We proceed as  
as before, changing variables $S_{i}=n_{i\uparrow}-n_{i\downarrow}$  
and $\tau_{i}=n_{i\uparrow}+n_{i\downarrow}$ and tracing out the  
dependence on the $\tau$'s.  We can then re-write the above expression  
as  
\begin{equation}   
\begin{split}   
\mathcal{G}_{i}^{s}(\omega)&=\int \, d h\frac{p^{\text{GS}}_i(h)}{\omega +(\mu+sh )/\hbar+i\delta}   
\end{split}   
\end{equation}   
with $p^{\text{GS}}_i(h)$ the density of local fields at site $i$, \textit{viz.}   
\begin{equation}   
\begin{split}   
p^{\text{GS}}_i(h)=\bracket{\delta\left[h- h_{i}(\bS)\right]}_{\mathcal{H}_{\text{GS}}}\,.   
\end{split}   
\end{equation}   
Defining the DoS as the imaginary part of the spectral density  
function averaged over all sites, spin orientation and over the  
disorder, denoting the latter by an overline  
$\overline{\,\,\,\bullet\,\,\,}$, using the identity  
$\frac{1}{x+i\delta}=\bP\frac{1}{x}-i\pi\delta(x)$ and shifting  
the energy levels $\epsilon=\hbar\omega+\mu$, we obtain finally  
\begin{equation}   
\begin{split}   
\rho^{\text{DoS}}(\epsilon)&=-\frac{1}{2\pi N}\sum_{s=\pm1}\sum_{i=1}^N\overline{\text{Im}\mathcal{G}_{i}^{s}(\omega)}\\  
&=\frac{1}{2}\sum_{s=\pm1}p^{\text{GS}}(s\epsilon)   
\label{eq:Dos_fields}   
\end{split}   
\end{equation}   
with definition   
\begin{equation}   
p^{\text{GS}}(h)\equiv\frac{1}{N}\sum_{i=1}^N\overline{p^{\text{GS}}_i(h)}   
\end{equation}   
If $\mathcal{J}_0=0$ then we have that the distribution of fields is
an even function, \textit{i.e.}
$p^{\text{GS}}(\epsilon)=p^{\text{GS}}(-\epsilon)$, and therefore the
expression \eqref{eq:Dos_fields} reveals that the DoS in the fermionic
Ising spin-glass model without Hubbard interaction is \textit{exactly}
the distribution of local fields in the corresponding classical
Ghatak-Sherrington model for \textit{any value} of the chemical
potential and temperature.
\subsection{Fermion Ising spin-glass model with Hubbard interaction}   
The fermionic Ising spin-glass model with Hubbard interaction was  
first studied in \cite{Oppermann2003}. Again, due to the term  
$|\bra{\textbf{n}}\widehat{a}_{i s}\ket{\textbf{m}}|^2$, in eq.  
\eqref{eq:greenfunction123} we can replace  
\begin{equation}   
\begin{split}   
\mathcal{H}(\textbf{m})-\mathcal{H}(\textbf{n})&=Un_{i\overline{s}}-sh_{i}(\bS)-\mu   
\end{split}   
\end{equation}   
to yield   
\begin{equation}   
\begin{split}   
\mathcal{G}_{i}^{s}(\omega)&=\sum_{\gamma=0,1}\int  
\frac{dh}{\omega+(\mu+sh-\gamma U )/\hbar+i\delta}\\  
&\times\mathcal{Z}^{-1}(\beta)\sum_{\textbf{n}}\delta_{n_{i\overline{s}},\gamma} 
e^{-\beta\mathcal{H}(\textbf{n})}\delta[h-h_{i}(\bsigma)]\,  
\end{split}   
\end{equation}   
where $\overline{s}\equiv-s$. We proceed as before and map to the GS  
model. This calculation is a bit more involved but fairly  
straightforward and after some algebra we arrive at  
\begin{equation}   
\begin{split}   
\mathcal{G}_{i}^{s}(\omega)&=\sum_{\gamma=0,1}\sum_{\tau=0,\pm1}a^\gamma_\tau(s)\int  
dh\frac{p^{\text{GS}}_i(\tau,h)}{\omega+(\mu+sh-\gamma U  
)/\hbar+i\delta}  
\label{eq:DoSHubbard}   
\end{split}   
\end{equation}   
where we have introduced the joint spin-field distribution at site $i$  
of the GS spin-glass  
\begin{equation}   
\begin{split}   
p^{\text{GS}}_i(\tau,h)=\bracket{\delta_{S_i,\tau}  
\delta\left[\epsilon-h_{i}(\bS)\right]}_{\mathcal{H}_{\text{GS}}}\,,  
\end{split}   
\end{equation}   
with    
\begin{equation}   
a^\gamma_\tau(s)=\delta_{\tau,0}\frac{\delta_{0,\gamma}+\delta_{1,\gamma}e^{-\beta  
U+2\beta\mu}}{1+e^{-\beta  
U+2\beta\mu}}+\delta_{0,\gamma}\delta_{\tau,s}+\delta_{1,\gamma}\delta_{\tau,\overline{s}}\,.  
\end{equation}   
From here we have that the DoS is given by  
\begin{equation}  
\begin{split}  
\rho^{\text{DoS}}(\epsilon)=\frac{1}{2}\sum_{s=\pm1}\sum_{\gamma=0,1}\sum_{\tau=0,\pm1}a^\gamma_\tau(s)p^{\text{GS}} 
[\tau,s(\gamma U-\epsilon)]  
\end{split}  
\end{equation}  
with  
\begin{equation}   
p^{\text{GS}}(\tau,h)\equiv\frac{1}{N}\sum_{i=1}^N\overline{p^{\text{GS}}_i(\tau,h)}   
\end{equation}   
In this case we have again an  
intimate relationship between the DoS in the fermonic Ising spin-glass  
model and the joint spin-field distribution of the classical GS  
spin-glass model.  
\section{physical mapping of DoS to field distributions}  
\label{sec:physical}  
In complementation of the formal mathematical mappings discussed  
above, in this section we describe how the above connection between  
the classical and quantum systems also appears naturally, based solely on  
physical arguments.  For the sake of simplicity we restrict the  
discussion to zero temperature and $\mathcal{J}_0=0$.  
  
Let us consider initially that $U=0$ and $\mu=0$, \textit{i.e.}  
half-filling.  In this case, the number of fermions $N_{\text{f}}$ is  
equal to the number of sites $N_{\text{sites}}$.  It is a reasonable  
ansatz, which we shall later show to be true, that in the ground state 
every site will  
carry a single fermion, whose spin can be either up or down. It is  
immediately clear that this is nothing but the usual classical SK  
model. Consequently the ground state of the fermionic Ising spin-glass  
model is the same as that of the classical SK model and the DoS of the  
former system is simply  
\begin{equation}   
\begin{split}   
\rho^{\text{DoS}}(\epsilon)=p^{\text{SK}}(-|{\epsilon}|)  
\end{split}.   
\end{equation}   
where   
\begin{equation}   
p^{\text{SK}}(h)=\frac{1}{N}\sum_{i=1}^N  
\bracket{\delta\left(h-\left|\sum_{j}\mathcal{J}_{ij}\sigma_j\right|\right)}_{\mathcal{H}_\text{SK}}.  
\end{equation}  
  
Next let us consider further the ansatz that each site is singly  
occupied.  Were a site to be unoccupied then clearly it would  
contribute no energy to the ground state. Neither would a doubly  
occupied site, since the two spins would both see the same effective  
field due to the other spins and they would contribute cancelling  
energies. However, we also know that in the ground state of the SK  
model all local fields are finite and that spins are oriented to yield  
negative energies. Consequently it follows that removing a fermion  
from one singly occupied site and depositing it on another, already  
favourably singly occupied, site incurs two energetic  
penalties. Furthermore, since the distribution of local fields in the  
SK model goes to zero at zero field and any single spin coupling  
strength scales as $N^{-1}$ the loss cannot be compensated by further  
re-adjustments on other sites.  
  
For $\mu \neq 0$ account must be taken of the fact that in the  
fermionic model some sites must be unoccupied for $\mu < 0$ or doubly  
occupied for $\mu > 0$. In both cases the system behaves energetically  
as though it were a diluted classical SK model with spins absent on  
the sites of either zero or double occupancy in the fermionic  
model. Furthermore, the location of these holes is chosen so as to  
minimise the total ground state energy; i.e. the system behaves as  
though one has an effective Hamiltonian  
\begin{equation}  
\begin{split}  
\label{eq:diluteSK}   
\mathcal{H}({\bsigma},{\textbf{n}})=&-\sum_{i<j=1}^N \mathcal{J}_{ij}  
\sigma_i{\sigma_j}{n_i}{n_j}-\widetilde{\mu}\sum_{i=1}^N{n_i}\\  
&\sigma =\pm 1\,,\quad n=0,1\,,  
\end{split}  
\end{equation}  
i.e. with two types of annealed variables, Ising spins (characterised 
by the {$\sigma$}) and `quasi-particles' (characterised by the {$n_i$} 
and not to be confused with the real fermions of number operator 
$\widehat{n}_{is}$). We shall refer to this system as the 
\textit{anneal-diluted SK model} (ADSK). The two chemical potentials, 
of eqs. \eqref{eq:diluteSK} and \eqref{eq:quantum_Hamiltonian}, are 
related by 
\begin{equation}   
\widetilde{\mu} = -|\mu|.   
\end{equation} 
The DoS is given by  
\begin{equation}   
\label{eq:rho_munonzero}  
\begin{split}   
\rho^{\text{DoS}}(\epsilon)=p^{\text{ADSK}}(-|{\epsilon}|),\quad|\epsilon| > |\tilde{\mu}|,  
\end{split}   
\end{equation} 
where $p^{\text{ADSK}}$ is defined analogously to $p^{\text{SK}}$ but 
with the sum over only the singly occupied sites and averaged over the 
ADSK Hamiltonian \eqref{eq:diluteSK}. 
  
It is tempting to think that the truncation of site occupation might  
modify the DoS of the $\mu=0$ case by simply moving the Fermi level of  
the $\rho^{\text{DoS}}(\epsilon)$ corresponding to the pure SK model  
so as to occupy only the lowest states up to $\tilde{\mu}$, but this  
does not take account of the loss of contribution to the fields of the  
unoccupied sites of \eqref{eq:diluteSK}; in fact, computer studies of  
the TAP equations have shown that $\rho^{\text{DoS}}(\epsilon)$ now  
goes to zero at $\epsilon = \tilde{\mu}$, in a manner at least  
qualitatively similar to what happens at $\epsilon=0$ for the case of  
$\mu=0$ \cite{Rehker1999}. A replica symmetric analysis of $\rho$  
close to $\epsilon = \tilde{\mu}$ also behaves analogously to the  
corresponding replica symmetric study for the undiluted SK model near  
$\epsilon = 0$; the full replica symmetry breaking calculation has not  
yet been done explicitly.  
  
For $\rho^{\text{DoS}}(\epsilon)$ with $|\epsilon| < |\tilde{\mu}|$ it  
is necessary to calculate the local field distribution $h_i=\sum_{j}  
\mathcal{J}_{ij}\sigma_j$ at sites of \eqref{eq:diluteSK} where there  
is no quasi-particle, so that such sites do not contribute to the  
total energy or the field or spin orientation at other sites. With  
this extension \eqref{eq:rho_munonzero} applies for all $\epsilon$.  
  
For $U > 0$ the mapping of \eqref{eq:diluteSK} continues to apply with 
$\tilde{\mu}$ appropriately chosen. Again if $\mu <U/2$ there are 
$(N_{\text{site}} - N_{\text{f}})$ unoccupied sites and for $\mu >U/2$ 
there are $(N_{\text{f}}-N_{\text{site}})$ doubly occupied fermion 
sites; consequently in both cases there are $|(N_{\text{f}} - 
N_{\text{site}})|$ sites without quasi-particles. $\widetilde{\mu}$ is 
given by \footnote{Note that $\widetilde{\mu}$ varies continuously with 
$N_f$, always non-positive and with $|\widetilde{\mu}|$ monotonically 
increasing with $|(N_f -N_\text{site})|$, whereas $\mu$ jumps 
discontinuously as $(N_f - N_\text{site})$ increases past 0 for $U\neq0$} 
\begin{equation}  
\widetilde{\mu}=\left\{\begin{array}{ll}    
 \mu,& \mu < U/2\\  
 U - \mu,& \mu > U/2  
\end{array}\right.   
\end{equation}  
and $\rho^{Dos}(\epsilon)$ is given by  
\begin{equation}   
\rho^{\text{DoS}}(\epsilon)=\left\{  
\begin{array}{ll}   
p^{\text{ADSK}}(-|{\epsilon}|)&\epsilon < 0\\  
0& 0 < \epsilon <U\\  
p^{\text{ADSK}}(-|{\epsilon}-U|)&  
\epsilon > 0.  
\end{array}\right.   
\label{eq:rho_Unonzero}  
\end{equation}  
 
These mappings may be related to those of the previous section by noting 
that Hamiltonian \eqref{eq:diluteSK} is also another way of writing the  
GS model of \eqref{eq:GS} with $\tilde{\mu} = D-T\ln2$ \cite{Crisanti2004}.

\section{Existence of a Pseudogap in Fermionic Ising spin glasses}  
\label{sec:pseudogap}   
Having demonstrated the mapping between DoS and distribution of  
fields, we can draw some conclusions and speculate about the nature of  
the pseudogap in the DoS at the Fermi energy \cite{Oppermann2003}.  First  
of all, let us notice that the nature of the pseudogap and of the  
strong corrections of different steps of replica symmetry breaking,  
cannot be but of classical origin. The \textit{strong corrections} to  
the DoS from RSB corrections found in fermionic Ising spin-glass  
models are \textit{common} in classical spin glasses.  
  
In particular is has been shown already for some time from
$\infty-$RSB calculations \cite{Sommers1984,Thomsen1986} and
Thouless-Anderson-Palmer equations \cite{Thouless1977}, that the field
distribution at zero temperature in the SK model vanishes with $h
\rightarrow 0$ as $p^{\text{SK}}(h)=a|h|$ (See also
\cite{OS2005}). This therefore predicts the existence of a pseudogap
for the fermionic Ising-spin glass model without Hubbard term and at
half-filling \cite{Oppermann1999a}.
  
The distribution of fields for fermonic SK models without Hubbard  
interaction for $\mu \neq 0$ was studied numerically using a TAP  
approach in \cite{Rehker1999} for $\mu > 0$, indicating that it seems  
to vanish, again quasi-linearly, at $h = \pm \mu $, implying that the  
DoS also presents pseudogaps.  
  
We might note also that the field distribution is very  
temperature-dependent and the pseudogap becomes filled in as the  
temperature rises; so therefore (and unusually) the density of  
fermionic states will mirror this strong temperature-dependence.\\  
   
\section{Conclusions}   
In this paper we have shown that the fermionic Ising spin-glass model
(with SK-like interactions) is mappable to the classical GS spin-glass
model not only at the level of the free energy but also the DoS, the
latter being given exactly by the local field distribution of the GS
model in the case without Hubbard interaction and obtainable from it
when a Hubbard term is present.  By using known results from spin
glass models we can show the existence of a pseudogap, from full RSB
and TAP approachs. It should be noted that the pseudogap and strong
corrections in the different steps of RSB are purely classical effects
and not due to quantum fluctuations. It would be interesting to see
how this picture changes when the Hamiltonian
$\widehat{\mathcal{H}}_{\text{rest}}$ is switched on and also when one
passes to a more realistic (but also more difficult to solve and
currently still controversial) model with short-range interactions.

Above we have concentrated on single-fermion Green functions and their
averages. A similar procedure to that outlined in Section III can be
applied to higher-order Green functions, mapping into higher order
field distributions, and for averages of products of Green functions.
  
Finally, we note that the GS model has a first-order phase transition
at a critical negative-valued $D_{c}(T)$ and beneath a tri-critical
temperature $T_3$, between a magnetic state ($|\bracket{S_i}|\neq0$)
and a non-magnetic ($|\bracket{S_i}|=0$) solution. This reflects in
the FISG to a critical $\mu_c$ for magnetic breakdown
\cite{Oppermann1999a}\footnote{More precisely, it leads to two
critical $\mu_c$, one positive, one negative.}.

\section*{Acknowledgments}   
The authors thank R. Oppermann for informative discussions. IPC  
also acknowledges R. Jack, B. Doyon, J. Chalker and F. Essler for  
 discussions. This work was supported by EPSRC under grant GR/R83712/01.

\bibliography{Isaac_DS_17_June}    

\begin{thebibliography}{20}
\expandafter\ifx\csname natexlab\endcsname\relax\def\natexlab#1{#1}\fi
\expandafter\ifx\csname bibnamefont\endcsname\relax
  \def\bibnamefont#1{#1}\fi
\expandafter\ifx\csname bibfnamefont\endcsname\relax
  \def\bibfnamefont#1{#1}\fi
\expandafter\ifx\csname citenamefont\endcsname\relax
  \def\citenamefont#1{#1}\fi
\expandafter\ifx\csname url\endcsname\relax
  \def\url#1{\texttt{#1}}\fi
\expandafter\ifx\csname urlprefix\endcsname\relax\def\urlprefix{URL }\fi
\providecommand{\bibinfo}[2]{#2}
\providecommand{\eprint}[2][]{\url{#2}}

\bibitem[{\citenamefont{Edwards and Anderson}(1975)}]{Edwards1975}
\bibinfo{author}{\bibfnamefont{S.~F.} \bibnamefont{Edwards}} \bibnamefont{and}
  \bibinfo{author}{\bibfnamefont{P.~W.} \bibnamefont{Anderson}},
  \bibinfo{journal}{J. Phys. F: Met. Phys.} \textbf{\bibinfo{volume}{5}},
  \bibinfo{pages}{965} (\bibinfo{year}{1975}).

\bibitem[{\citenamefont{Sherrington and Kirkpatrick}(1975)}]{Sherrington1975}
\bibinfo{author}{\bibfnamefont{D.}~\bibnamefont{Sherrington}} \bibnamefont{and}
  \bibinfo{author}{\bibfnamefont{S.}~\bibnamefont{Kirkpatrick}},
  \bibinfo{journal}{Phys. Rev. Lett.} \textbf{\bibinfo{volume}{35}},
  \bibinfo{pages}{1792} (\bibinfo{year}{1975}).

\bibitem[{\citenamefont{Sherrington and Mihill}(1974)}]{Sherrington1974}
\bibinfo{author}{\bibfnamefont{D.}~\bibnamefont{Sherrington}} \bibnamefont{and}
  \bibinfo{author}{\bibfnamefont{K.}~\bibnamefont{Mihill}},
  \bibinfo{journal}{J. Phys. (Paris) C4-199} \textbf{\bibinfo{volume}{35}}
  (\bibinfo{year}{1974}).

\bibitem[{\citenamefont{Hertz}(1979)}]{Hertz1979}
\bibinfo{author}{\bibfnamefont{J.~A.} \bibnamefont{Hertz}},
  \bibinfo{journal}{Phys. Rev. B} \textbf{\bibinfo{volume}{19}},
  \bibinfo{pages}{4796} (\bibinfo{year}{1979}).

\bibitem[{\citenamefont{Oppermann and M\"uller-Groeling}(1993)}]{Oppermann1993}
\bibinfo{author}{\bibfnamefont{R.}~\bibnamefont{Oppermann}} \bibnamefont{and}
  \bibinfo{author}{\bibfnamefont{A.}~\bibnamefont{M\"uller-Groeling}},
  \bibinfo{journal}{Nuclear Physics B,} \textbf{\bibinfo{volume}{401}},
  \bibinfo{pages}{507} (\bibinfo{year}{1993}).

\bibitem[{\citenamefont{Rosenow and Oppermann}(1996)}]{Rosenow1996}
\bibinfo{author}{\bibfnamefont{B.}~\bibnamefont{Rosenow}} \bibnamefont{and}
  \bibinfo{author}{\bibfnamefont{R.}~\bibnamefont{Oppermann}},
  \bibinfo{journal}{Phys. Rev. Lett.} \textbf{\bibinfo{volume}{77}},
  \bibinfo{pages}{1608} (\bibinfo{year}{1996}).

\bibitem[{\citenamefont{Oppermann and
  Rosenow}(1998{\natexlab{a}})}]{Oppermann1998}
\bibinfo{author}{\bibfnamefont{R.}~\bibnamefont{Oppermann}} \bibnamefont{and}
  \bibinfo{author}{\bibfnamefont{B.}~\bibnamefont{Rosenow}},
  \bibinfo{journal}{Europhys. Lett} \textbf{\bibinfo{volume}{41 (5)}},
  \bibinfo{pages}{525} (\bibinfo{year}{1998}{\natexlab{a}}).

\bibitem[{\citenamefont{Oppermann and
  Rosenow}(1998{\natexlab{b}})}]{Oppermann1998a}
\bibinfo{author}{\bibfnamefont{R.}~\bibnamefont{Oppermann}} \bibnamefont{and}
  \bibinfo{author}{\bibfnamefont{B.}~\bibnamefont{Rosenow}},
  \bibinfo{journal}{Phys. Rev. Lett.} \textbf{\bibinfo{volume}{80}},
  \bibinfo{pages}{4767} (\bibinfo{year}{1998}{\natexlab{b}}).

\bibitem[{\citenamefont{Oppermann and Rosenow}(1999)}]{Oppermann1999}
\bibinfo{author}{\bibfnamefont{R.}~\bibnamefont{Oppermann}} \bibnamefont{and}
  \bibinfo{author}{\bibfnamefont{B.}~\bibnamefont{Rosenow}},
  \bibinfo{journal}{Phys. Rev. B} \textbf{\bibinfo{volume}{60}},
  \bibinfo{pages}{10325} (\bibinfo{year}{1999}).

\bibitem[{\citenamefont{Rehker and Oppermann}(1999)}]{Rehker1999}
\bibinfo{author}{\bibfnamefont{M.}~\bibnamefont{Rehker}} \bibnamefont{and}
  \bibinfo{author}{\bibfnamefont{R.}~\bibnamefont{Oppermann}},
  \bibinfo{journal}{J. Phys. C} \textbf{\bibinfo{volume}{11}},
  \bibinfo{pages}{1537} (\bibinfo{year}{1999}).

\bibitem[{\citenamefont{Feldmann and
  Oppermann}(2000{\natexlab{a}})}]{Feldmann2000a}
\bibinfo{author}{\bibfnamefont{H.}~\bibnamefont{Feldmann}} \bibnamefont{and}
  \bibinfo{author}{\bibfnamefont{R.}~\bibnamefont{Oppermann}},
  \bibinfo{journal}{J. Phys. A: Math. Gen.} \textbf{\bibinfo{volume}{33}},
  \bibinfo{pages}{1325} (\bibinfo{year}{2000}{\natexlab{a}}).

\bibitem[{\citenamefont{Feldmann and
  Oppermann}(2000{\natexlab{b}})}]{Feldmann2000}
\bibinfo{author}{\bibfnamefont{H.}~\bibnamefont{Feldmann}} \bibnamefont{and}
  \bibinfo{author}{\bibfnamefont{R.}~\bibnamefont{Oppermann}},
  \bibinfo{journal}{Phys. Rev. B} \textbf{\bibinfo{volume}{62}},
  \bibinfo{pages}{9030} (\bibinfo{year}{2000}{\natexlab{b}}).

\bibitem[{\citenamefont{Oppermann and Sherrington}(2003)}]{Oppermann2003}
\bibinfo{author}{\bibfnamefont{R.}~\bibnamefont{Oppermann}} \bibnamefont{and}
  \bibinfo{author}{\bibfnamefont{D.}~\bibnamefont{Sherrington}},
  \bibinfo{journal}{Phys. Rev. B} \textbf{\bibinfo{volume}{67}},
  \bibinfo{pages}{245111} (\bibinfo{year}{2003}).

\bibitem[{\citenamefont{Oppermann and Feldmann}(1999)}]{Oppermann1999a}
\bibinfo{author}{\bibfnamefont{R.}~\bibnamefont{Oppermann}} \bibnamefont{and}
  \bibinfo{author}{\bibfnamefont{H.}~\bibnamefont{Feldmann}},
  \bibinfo{journal}{J Phys IV} \textbf{\bibinfo{volume}{9 (P10)}},
  \bibinfo{pages}{37} (\bibinfo{year}{1999}).

\bibitem[{\citenamefont{Ghatak and Sherrington}(1977)}]{Ghatak1977}
\bibinfo{author}{\bibfnamefont{S.~K.} \bibnamefont{Ghatak}} \bibnamefont{and}
  \bibinfo{author}{\bibfnamefont{D.}~\bibnamefont{Sherrington}},
  \bibinfo{journal}{J. Phys. C} \textbf{\bibinfo{volume}{10}},
  \bibinfo{pages}{3149} (\bibinfo{year}{1977}).

\bibitem[{\citenamefont{Crisanti and Leuzzi}(2004)}]{Crisanti2004}
\bibinfo{author}{\bibfnamefont{A.}~\bibnamefont{Crisanti}} \bibnamefont{and}
  \bibinfo{author}{\bibfnamefont{L.}~\bibnamefont{Leuzzi}},
  \bibinfo{journal}{Phys. Rev. B} \textbf{\bibinfo{volume}{70}},
  \bibinfo{pages}{014409} (\bibinfo{year}{2004}).

\bibitem[{\citenamefont{Sommers and Dupont}(1984)}]{Sommers1984}
\bibinfo{author}{\bibfnamefont{H.~J.} \bibnamefont{Sommers}} \bibnamefont{and}
  \bibinfo{author}{\bibfnamefont{W.}~\bibnamefont{Dupont}},
  \bibinfo{journal}{J. Phys. C: Solid State Phys.}
  \textbf{\bibinfo{volume}{17}}, \bibinfo{pages}{5785} (\bibinfo{year}{1984}).

\bibitem[{\citenamefont{Thomsen et~al.}(1986)\citenamefont{Thomsen, Thorpe,
  Choy, Sherrington, and Sommers}}]{Thomsen1986}
\bibinfo{author}{\bibfnamefont{M.}~\bibnamefont{Thomsen}},
  \bibinfo{author}{\bibfnamefont{M.~F.} \bibnamefont{Thorpe}},
  \bibinfo{author}{\bibfnamefont{T.~C.} \bibnamefont{Choy}},
  \bibinfo{author}{\bibfnamefont{D.}~\bibnamefont{Sherrington}},
  \bibnamefont{and} \bibinfo{author}{\bibfnamefont{H.-J.}
  \bibnamefont{Sommers}}, \bibinfo{journal}{Phys. Rev. B}
  \textbf{\bibinfo{volume}{33}}, \bibinfo{pages}{1931} (\bibinfo{year}{1986}).

\bibitem[{\citenamefont{Thouless et~al.}(1977)\citenamefont{Thouless, Anderson,
  and Palmer}}]{Thouless1977}
\bibinfo{author}{\bibfnamefont{D.~J.} \bibnamefont{Thouless}},
  \bibinfo{author}{\bibfnamefont{P.~W.} \bibnamefont{Anderson}},
  \bibnamefont{and} \bibinfo{author}{\bibfnamefont{R.~G.}
  \bibnamefont{Palmer}}, \bibinfo{journal}{Philosophical Magazine}
  \textbf{\bibinfo{volume}{35}}, \bibinfo{pages}{593} (\bibinfo{year}{1977}).

\bibitem[{\citenamefont{Oppermann and Sherrington}(2005)}]{OS2005}
\bibinfo{author}{\bibfnamefont{R.}~\bibnamefont{Oppermann}} \bibnamefont{and}
  \bibinfo{author}{\bibfnamefont{D.}~\bibnamefont{Sherrington}},
  \bibinfo{journal}{to be published}  (\bibinfo{year}{2005}).

\end{thebibliography}
\end{document}